\documentstyle[aps,epsf]{revtex}  % DO NOT DELETE THIS LINE 

\global\firstfigfalse  % kludge to bypass foolishness of revtex to place
\global\firsttabfalse  % figures and tables at end of paper

\def\der    {\partial }
\def\bra    {\langle }
\def\ket    {\rangle }
\def\ra     {\rightarrow }
\def\A#1    {{\cal A}_{#1} }
\def\Afbp#1 {{\mathrm A_{FB}^{#1,0}} }
\def\Afb#1  {{\mathrm A_{FB}^{#1}} }
\def\Alr    {{\mathrm A_{LR}} }
\def\voa#1  {\frac{v_{#1}}{a_{#1}} }
\def\sW     {\sin^2\theta_W }
\def\sWe#1  {\sin^2\theta_{W,eff}^{#1} }
\def\MZ     {M_Z } 

\def\ff#1   {{\mathrm #1\bar{#1}} }
\def\B#1    {{\mathrm B}^{#1} }
\def\D#1    {{\mathrm D}^{#1} }
\def\K#1    {{\mathrm K}^{#1} }
\def\R#1    {{\mathrm R_{#1}} }
\def\Rp#1   {{\mathrm R_{#1}^{0}} }
\def\eff#1  {\varepsilon_{#1}}
\begin{document}        %  DO NOT DELETE OR CHANGE THIS LINE

\baselineskip 14pt
\title{Heavy quark asymmetries with DELPHI}
\author{Ernesto Migliore}
\address{CERN, CH-1211 Geneva 23, Switzerland}
\maketitle              % Creates the title area, Do Not Remove

\begin{abstract}        % Do Not Delete this line
The measurements of the forward-backward asymmetry in $\mathrm{Z} \ra \ff{c} $ 
and $\mathrm{Z} \ra \ff{b} $ decays are among the most precise determinations
of $\sWe{\ell } $. In this paper the results obtained by the DELPHI
experiment at LEP with three different analyses are reviewed together
with the impact of the combined LEP result on the global Electroweak fit.
\
\end{abstract}          % Do Not Delete this line

\section{Introduction}               % Introduction goes below.

As a consequence of the parity violating couplings of the Z boson to the 
fermions, in electron positron annihilation at the Z mass 
fermions are more likely produced in the forward direction, with respect the 
incoming electron, than in the backward.
In the Electroweak Standard Model the asymmetry at the Z pole~\footnote{
Throughout this paper pseudo-observables defined at the Z pole are indicated 
with an apex 0 while values measured at the Z peak ($\sqrt{s} = 91.26$ GeV)
without.} is expressed as:

\begin{eqnarray*}
\Afbp{f} =\frac{3}{4} \A{e} \A{f}
\end{eqnarray*}

\noindent $\A{e} $ and $\A{f} $ are functions of the ratio $x_f$ of vector 
$v_f$ and axial $a_f$ couplings of the Z boson to the fermions:

\begin{eqnarray*}
\A{f} = \frac{2 x_f}{1+x_f^2}
\end{eqnarray*}
 
\noindent The ratio depends on the quantum numbers of the fermion, the charge 
$Q_f$ and the weak isospin $I_{3,f}$, and on the fundamental parameter of the 
theory the electroweak mixing angle $\sW $: $x_f = 1-4|Q_f| \sW $.

\begin{figure}[ht]      % in second brace, h=here, t=top, b=bottom      
\centerline{\epsfxsize 2.1 truein \epsfbox{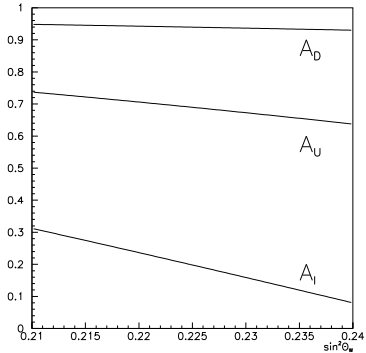}}   
\vskip -.2 cm
\caption[]{
\label{fig:sW}
\small Dependence of Born level $\A{f} $ on $\sW $ for leptons, up-type quarks
and down-type quarks.}
\end{figure}

\noindent The dependence of the Born level $\A{f} $ on $\sW $ for different 
fermion 
species is shown in fig~\ref{fig:sW}: For $\sW \simeq 0.23$ the asymmetry for 
$\ff{q} $ final states is from 3 to 5 times larger than for 
leptonic final states.

As among all the $\ff{q} $ final states the decays of the Z boson into 
$\ff{c} $ and $\ff{b} $ pairs are the most easily distinguished, the 
experimental effort 
was concentrated on the measurement of $\Afb{c} $ and $\Afb{b} $.

Most of the electroweak radiative corrections can be accounted for 
by introducing in the Born level equations an effective electroweak mixing 
angle $\sWe{f} $ defined for each fermion family. 
On the contrary the sensitivity of the quark asymmetry to the final state 
couplings is heavily suppressed. This can be understood from the
the dependence of the sensitivity on $Q_f$ and $x_f$:

\begin{eqnarray*}
\frac{1}{\A{f} } \frac{\der \A{f} }{\sWe{f} } = 
4 |Q_f| \frac{1-x_f^2}{x_f (1+x_f^2)}
\end{eqnarray*}

\noindent The final state $c$ and the $b$ quarks have a smaller charge and 
a higher value of $x_f$ compared with the initial state leptons. 
The sensitivity to the quark couplings is reduced of about a factor 50 for 
the $b$ and 5 for the $c$ asymmetry.

The main motivation for  a precise measurement of the asymmetry is that $m_H$, 
the mass of the still undiscovered Higgs boson, enters in the definition of 
$\sWe{\ell} $. Table~\ref{Intro:tab1} shows that, even if the 
dependence is only logarithmic, the measurement of $\Afbp{b} $ sets currently 
the most stringent limits on $m_H$: A 4 \% accuracy in the determination 
corresponds to values of the mass of the Higgs boson in a range from 70 
GeV/$c^2$ to 1000 GeV/$c^2$. On the contrary the ratios of partial widths of 
the Z boson into $\ff{c} $ and $\ff{b} $ pairs are, in practice, not sensitive 
to $m_H$ at all. The current precision on $\Afbp{b} $ is smaller than this 
value and the prediction on the Higgs boson mass is spoiled mainly by the 
uncertainty on $\alpha_{QED}(\MZ^2)$.
\enlargethispage*{100cm}
\begin{table}
 \caption{ Dependence of the electroweak pseudo observables in the heavy 
 flavours sector on radiative correction (from~\protect\cite{Moenig} ). }
 \label{Intro:tab1}
 \begin{tabular}{ccccc}
   & $m_t = 175 \, \mathrm{GeV}/c^2$     & $\pm 6 \, \mathrm{GeV}/c^2$           & & \\  
   & $m_H = 300 \, \mathrm{GeV}/c^2$     & & $ ^{+700}_{-230} \, \mathrm{GeV}/c^2$ & \\  
   & $1/\alpha_{QED}(\MZ^2) = 128.896$ & & & $\pm 0.090$                       \\  
 \tableline 
 $\Afbp{b} $ & 0.0998& $\pm 0.0010 $ & $^{-0.0037}_{+0.0042}$ & $\mp 0.0013$ \\
 $\Afbp{c} $ & 0.0711& $\pm 0.0008 $ & $^{-0.0028}_{+0.0032}$ & $\mp 0.0010$ \\
 $\Rp{b}   $ & 0.2158& $\mp 0.0002 $ & 0                      & 0            \\
 $\Rp{c}   $ & 0.1722& $\pm 0.0001 $ & 0                      & 0            \\
 \end{tabular}
 \end{table}

\section{Experimental techniques.}
All the analyses of the asymmetry follow three steps: The tag of the
flavour of the decay of the Z boson, the determination of the
direction of the primary quark and the separation of quark and
antiquark hemispheres. The analyses described~\cite{delphi-papers} 
differ on how the tag is performed: Exploiting the long flight distance of 
the heavy flavoured hadron, the presence of a high momentum lepton or finally a
reconstructed $\D{} $ meson in the final state~\footnote{If not explicitly  
mentioned charge conjugate states are implicitly included.}. This
choice determines also how the assignment of
quark-antiquark hemispheres is done: Using
the charge flow in the event or the charge of the identified
lepton or the charge state of the $\D{} $ meson. Common to the
analyses is the determination of the direction of the initial $\ff{q}
$ pair done with the thrust axis of the event
conventionally oriented to form an angle $\theta_T < 90^\circ$ with the
direction of the incoming electron beam. The observed asymmetry is
finally extracted from a $\chi^2$ fit to the differential
asymmetry $\frac{N_i^+ - N_i^-}{N_i^+ + N_i^-}$ where $N_i^+(N_i^-)$
is the number of forward (backward) events in bin $i$ of $\cos\theta_T$.

\subsection{$\Afbp{b} $ using jet charge technique.}

Methods exploiting the long fly distance of \mbox{$b$-hadrons}
($\gamma \beta c \tau_b \sim 2$ mm) and the accurate resolution on
secondary vertex reconstruction of silicon vertex detectors provide
the most efficient way to select a sample of $\ff{b} $ 
events. In order to maximize the efficiency versus purity performance 
the $b$-tag algorithm used in the DELPHI experiment 
combines the informations from the reconstructed secondary vertex 
(its effective mass, the rapidity of the tracks associated to it and the 
fraction of the jet energy carried by them) with the lifetime 
information.
\pagebreak
\begin{figure}[T]      % in second brace, h=here, t=top, b=bottom      
\centerline{\epsfysize 1.6 truein \epsfbox{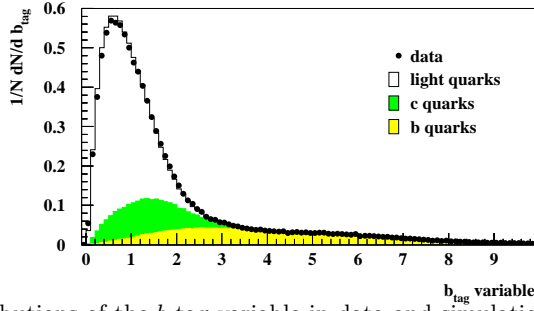}}   
\vskip -.2 cm
\caption[]{
\label{fig:JC1}
\small Distributions of the $b$-tag variable in data and simulation for 1994 
data.}
\end{figure}
In the measurement of $\Afbp{b}  $ the fraction $P_b(cut)$ of $\ff{b} $ events
in the sample after a cut in the $b$-tag variable is determined 
from the data themselves using the relation:

\begin{eqnarray*}
P_b(cut) = 
\frac{F(cut)-\R{c} \times \eff{c} (cut)-(1-\R{c} -\R{b} ) \times \eff{uds} 
(cut)}{F(cut)}
\end{eqnarray*}

\noindent where $F(cut)$ is the fraction of the data after the cut, 
$\eff{uds} (cut)$ and $\eff{c} (cut)$ the efficiencies for light ($q=u,d,s)$
and charm quark taken  from the simulation and 
$\R{c} $ and $\R{b} $ the partial widths of the Z into $\ff{c} $ and $\ff{b} $
 pairs from the Standard Model. 
\noindent Fig.~\ref{fig:JC1} shows the distribution of the $b$-tag variable in 
data and in simulation. The value of the tag chosen for the analysis 
corresponds to $\eff{b} = 75 \%$ and $P_b = 92 \%$.

The separation of $b$ from $\bar{b}$ quark relies on the hemisphere charge 
$Q_{hem}$ defined as the sum of the charges $q_i$ of the tracks in each 
hemisphere, 
as defined by the thrust axis $\vec{T}$, weighted by the projection of
their momenta $\vec{p_i}$ along $\vec{T}$ itself, to some power $\kappa$:

\begin{eqnarray*}
Q_{hem} = \frac{\sum_i q_i |\vec{p_i} \cdot \vec{T}|^\kappa}
               {\sum_i |\vec{p_i} \cdot \vec{T}|^\kappa}
\end{eqnarray*}

\noindent This estimator is based on the fact that, due to the electric charge 
conservation, the particles produced in the hadronization of the
primary quark retain some information of its charge. 
In the analysis done by DELPHI $\kappa = 0.8$. This choice minimizes 
the total statistical and systematic error.

In each event the total charge $Q_{TOT} = Q_F+Q_B$ and the charge flow 
$Q_{FB}=Q_F-Q_B$ are measured. 
A part from reinteractions $\bra Q_{TOT} \ket \simeq 0$, while the average 
charge flow is directly related to the asymmetries:

\begin{eqnarray}
\bra Q_{FB} \ket = \sum P_q \eta_q \delta_q \Afb{q}
\label{eq:JC1}
\end{eqnarray}

\begin{figure}[b]      % in second brace, h=here, t=top, b=bottom      
\centerline{
\begin{tabular}{cc}
\epsfysize 1.75 truein \epsfbox{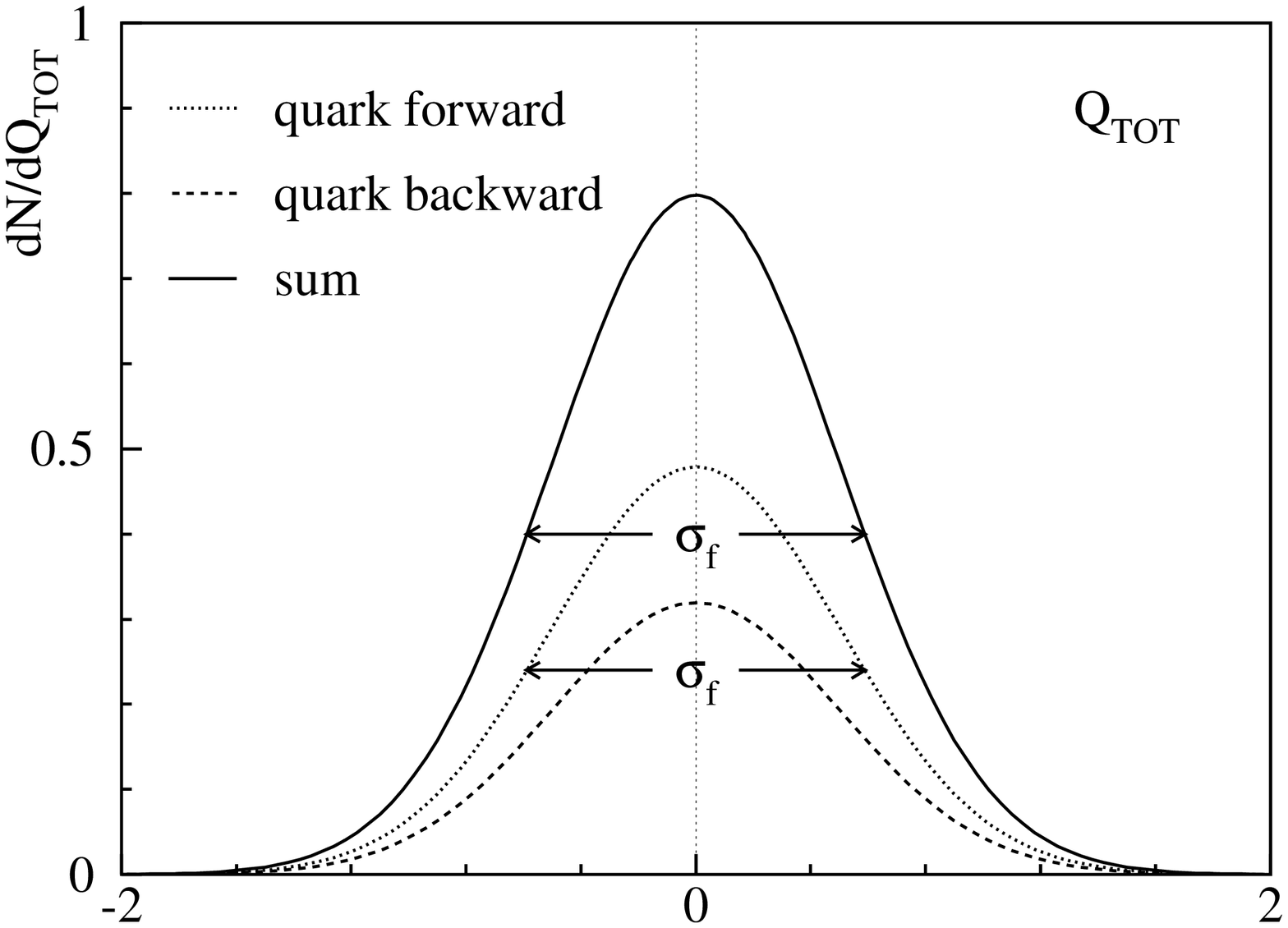}   & \epsfysize 1.75 truein \epsfbox{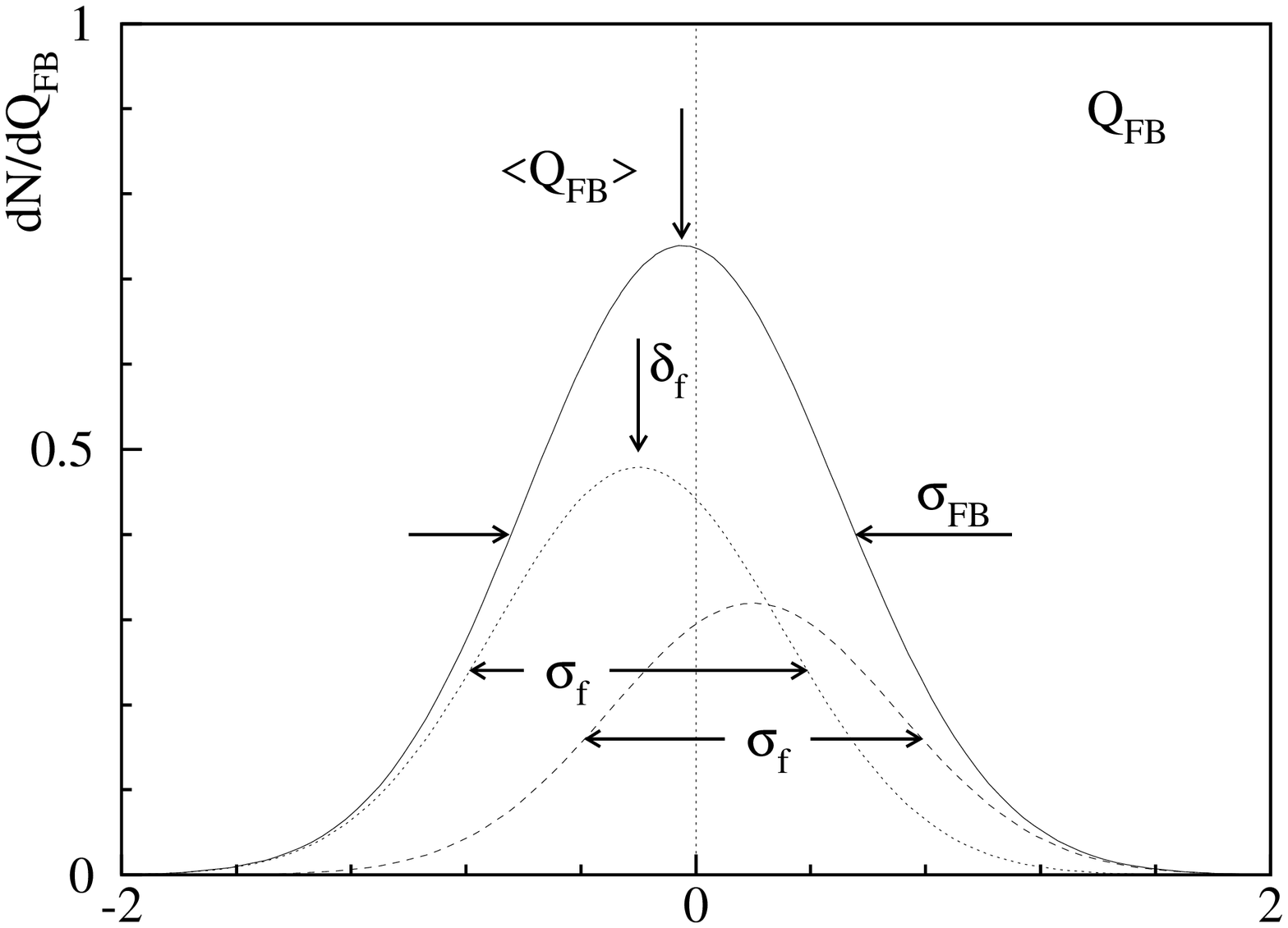}   \\
\end{tabular}}
\vskip -.2 cm
\caption[]{
\label{fig:JC2}
\small Sketch of the principle of the $\bra Q_{FB} \ket$ and the 
  $\delta_f$ measurement for a single (down type) flavour f.}
\end{figure}

\noindent $P_q$ is the fraction of events of flavour $q$ in  the sample, 
$\eta_q$ is an acceptance correction factor
and $\delta_q =\bra Q_q -Q_{\bar{q}} \ket $ is the charge separation: 
It would be twice the charge of the quark if quarks would be observed directly.
As for $P_b$, the charge separation for $b$ quark is measured directly in the 
data.
The principle of the method is sketched in fig.~\ref{fig:JC2}: For a pure $b$ 
sample the charge separation leads to an increase of the spread $\sigma_{FB}$ 
of the distribution of $Q_{FB}$ compared to $\sigma_{TOT}$ from the distribution of 
$Q_{TOT}$ so that $\delta_b^2 \simeq \sigma_{FB}^2-\sigma_{TOT}^2$.
As for the sample composition $\delta_{u,d,s,c}$ are taken from the 
simulation, carefully tuned in order to reproduce the measured 
distributions of hadronic event shapes and charged particle inclusive 
quantities.

\begin{figure}[b]      % in second brace, h=here, t=top, b=bottom      
\centerline{\epsfysize 1.4 truein \epsfbox{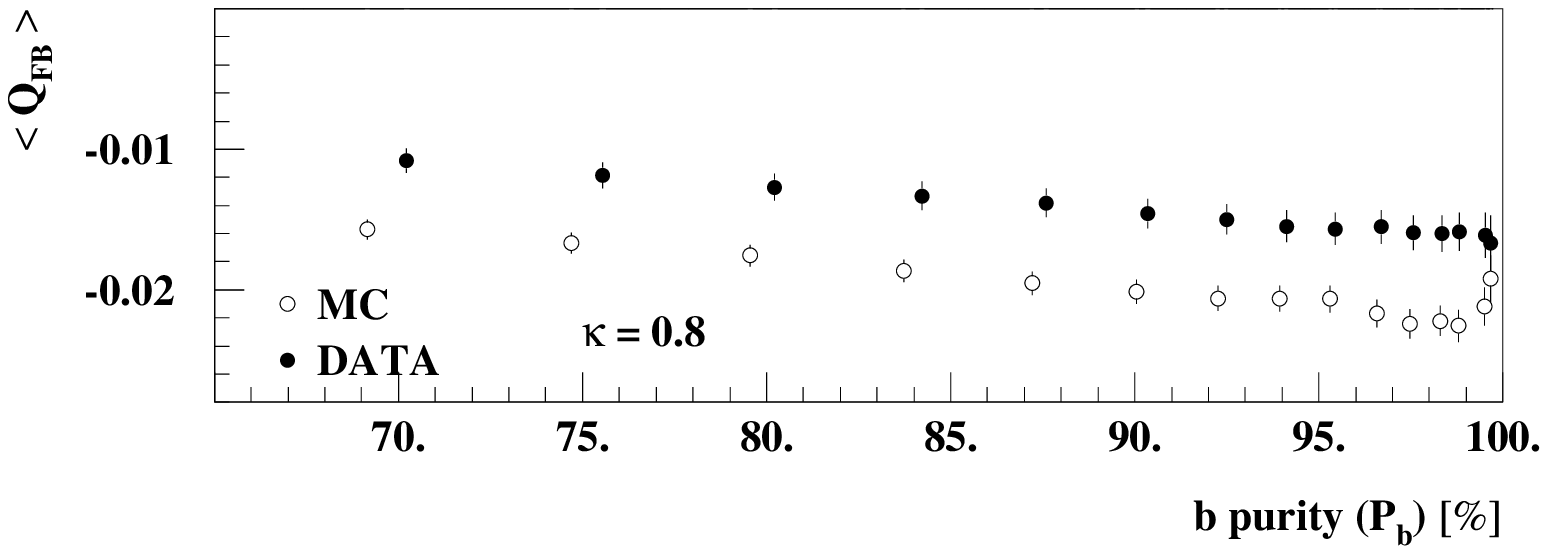}}   
%\vskip -.2 cm
\caption[]{
\label{fig:JC3}
\small $\bra Q_{FB} \ket$ as obtained from data and simulation for 
$\kappa=0.8$ and for 1994 data.}
\end{figure}

\noindent The observable $\bra Q_{FB} \ket$ in data and simulation as a function of 
increasing $b$-purity of the sample is shown in fig~\ref{fig:JC3}: The 
observed difference is due to different values of both $\delta_b$ and the 
input asymmetry between data and simulation.

To exploit the angular dependence of the asymmetry itself, $\Afb{b} $
is extracted from a $\chi^2$ fit of equation~\ref{eq:JC1} to
$\bra Q_{FB} \ket$ distribution in 4 bins of $\cos \theta_T$, the
acceptance being limited to $\theta_T>35^\circ$
by the actual angular coverage of the vertex detector.

The results for the Z peak asymmetry for 1992-95 sample
is $\Afb{b} = 0.0982 \pm 0.0047$ (stat).
The main contributions to the systematic error are listed in the top part of 
table~\ref{tab:syst}. The largest contribution ($\pm 0.0011$) comes from the 
hemisphere-hemisphere charge correlation which is due to charge conservation, 
the common thrust axis and the particle crossovers between hemispheres. 
This correlation is checked between data and simulation 
by  means of the variable $H$ defined such that differences in the description 
of the secondary interactions mostly cancel out:
$H=\bra Q_F Q_B\ket - \bra Q_{TOT}\ket^2/4$.
The systematic error accounts for 20 \%  discrepancy between data and 
simulation found in the distribution of $H$.

Effects related to the description of the physical processes which could 
effect the $b$-tag performance (gluon splitting into $\ff{c} $ and $\ff{b} $,
 $\K{0} $ and $\Lambda$ content in light quark events, lifetimes, spectra and 
production fraction of $\D{} $ mesons) are smaller than $\pm 0.0003$ each.

\enlargethispage*{100cm}
 \begin{table}[t]
 \caption{Main contributions, in units of $10^{-3}$, to the systematic error to $\Afb{c} $ 
  and $\Afb{b} $ for Z peak data.}
 \begin{tabular}{lcdcd}
                     &                       & $\delta \Afb{c} $&                           & $\delta \Afb{b} $\\
 \tableline  
  Jet Charge tag     &                              &           & $\Rp{c} \pm 5 \%$                 & $\pm$ 0.4 \\ 
                     &                              &           & $H$ charge correlation            & $\pm$ 1.1 \\
                     &                              &           & Detector resolution               & $\pm$ 0.7 \\
 \tableline  
                     & total                        &           &                                   & $\pm$ 1.6 \\
 \tableline
  leptons tag        & $c$ decay model              & $\mp$ 1.8 &                                   & $\pm$ 1.4 \\ 
                     & $b \ra \bar{c} \ra \ell $    & $\pm$ 1.9 & $p_\perp$ and jet reconstruction  & $\pm$ 1.3 \\
                     & bgd asymmetry                & $\pm$ 4.7 & $\B{0} \bar{\B{0} } $ mixing      & $\pm$ 1.1 \\
 \tableline  
                     & total                        & $\pm$ 6.3 &                                   & $\pm$ 2.7 \\
 \tableline
  $\D{} $ mesons tag & MC stat                      & $\pm$ 2.5 &                                   & $\pm$ 3.5 \\
                     & fit method                   & $\pm$ 1.7 &                                   & $\mp$ 2.8 \\
                     & bgd asymmetry ($b, c$ quark) & $\mp$ 1.3 &                                   & $\mp$ 3.6 \\
                     &                              &           & $\chi_{eff}$                      & $\pm$ 5.8 \\
 \tableline  
                     & total                        & $\pm$ 3.5 &                                   & $\mp$ 8.5 \\
 \end{tabular}
 \label{tab:syst}
 \end{table}

\subsection{$\Afbp{c} $ and $\Afbp{b} $ using prompt leptons.}

Electrons and muons are produced in the decay of heavy flavoured hadrons 
mainly by 3 
processes with a branching ratio of about 10 \% each: 
\begin{itemize}
\item primary semileptonic $b$ decays, $b \ra \ell^-$
\item weak cascades of $b$ hadrons, $\bar{b} \ra \bar{c} \ra \ell^-$
\item primary semileptonic $c$ decays, $\bar{c} \ra \ell^-$
\end{itemize}
\pagebreak

These leptons can be used to select on  statistical basis
$\ff{c} $ or $\ff{b} $ events as they have momentum $(p)$ spectrum harder 
than fragmentation particles, $\bra x_E \ket_c \sim 0.5$ and $\bra x_E \ket_b \sim 0.7$, 
and high transverse momentum $(p_\perp)$ with respect the direction of
the jet~\footnote{The axis of the jet is defined excluding 
the lepton momentum.}, because of the high mass of the parent hadron. 

The correlation between the charge of the lepton and the one of
the quark is exploited to determine the quark hemisphere.
The correlation is different in the three classes, resulting in a
dilution of the measured asymmetry:

\begin{eqnarray}
  \Afb{obs} = (1-2\overline{\chi}) (f_b-f_{bc}) \Afb{b} - f_c \Afb{c}
               + f_{bgd} \Afb{bgd}
\label{eq:lept1}
\end{eqnarray}

\noindent where the $f_i$'s are the composition of the sample, 
$(1-2\overline{\chi})$ is the dilution factor due to the $\B{0} \bar{\B{0} } $
mixing and the last term accounts for a non zero asymmetry
of the misidentified hadrons (mainly punch through hadrons and photon 
conversions) and non prompt leptons from $\K{} $ and $\pi$ decays.

DELPHI has recently introduced in the analysis of the data collected in 
1994-95 the use of lifetime based $b$-tag variable which is used to remove 
the decays of the Z into light quarks final states. 
This allows an estimation of the background level less dependent on the 
momentum of the lepton compared to the previous procedure. 
This consisted first in the selection of a high $(p,p_\perp)$ region, 
highly pure in the lepton content, which fixed the lepton identification 
efficiency and then 
in the evaluation of the background level in a low $(p,p_\perp)$ region.
Moreover the use of the $b$-tag enriches the sample in $\ff{c} $ events 
increasing the  sensitivity of the measurement to $\Afb{c} $. 
On the contrary the reduction of the statistical error for the $b$ asymmetry 
is smaller.
This is because the useful sample is limited to the region of $p_T >$ 1 
GeV/$c$ as the lower $p_T$ region is equally populated of primary $b$ and 
cascade leptons giving opposite 
sign contributions to the observed asymmetry: The $b$-tag variable has little
effect in separating these two classes of events.

The results for the Z peak asymmetry for 1991-95 statistics
are $\Afb{c} = 0.0770 \pm 0.0113$ (stat) and $\Afb{b} = 0.0998 \pm 0.0065$ 
(stat). 
The main contributions to the systematic error are listed in the central part 
of table~\ref{tab:syst}. The various contributions can be separated into two 
categories:
The terms arising from the model actually used to simulate the semileptonic 
decay processes and the ones related to the description of detector effects. 
In case of the $b$ asymmetry they both amount at about $\pm 0.002$ the largest 
effect coming from the description of the $c$ decay ($\pm 0.0014$).

\subsection{$\Afbp{c} $ and $\Afbp{b} $ using reconstructed $\D{} $ mesons.}

A reconstructed $\D{} $ meson uniquely tags a hadronic decay of the Z boson 
into a heavy flavour $\ff{q} $ pair as it can be produced only in the 
hadronization of a primarily produced $\ff{c} $ pair or in the decay of a 
$b$-hadron in a $\ff{b} $ event.
The charge state of the reconstructed $\D{} $ meson is correlated 
with the charge of the parent quark. 
Therefore the values of $\Afbp{c} $ and $\Afbp{b} $ are extracted form a 
$\chi^2$ fit to the distribution of $\cos \theta_T$ in $\D{} $ meson 
events.

\noindent The analysis done by DELPHI is based on the investigation of 8 decays channels:

\begin{tabular}{lll}
 $\D{*+} $ & $\ra \D{0}  \pi^+_{sl}$ &                    \\
          && $\ra (\K{-} \pi^+) \pi^+_{sl}$               \\ 
          && $\ra (\K{-} \pi^+ \pi^- \pi^+) \pi^+_{sl}$   \\
          && $\ra (\K{-} \pi^+ (\pi^0) ) \pi^+_{sl}$      \\
          && $\ra (\K{-} \mu^+ \bar{\nu_\mu} ) \pi^+_{sl}$\\
          && $\ra (\K{-}  e^+ \bar{\nu_e} ) \pi^+_{sl}$   \\

$\D{0} $ & $\ra \K{-} \pi^+$          & \\
$\D{0} $ & $\ra \K{-} \pi^+ (\pi^0)$  & \\

$\D{+} $ & $\ra (\K{-} \pi^+) \pi^+$  & \\
\end{tabular}

\noindent 
The first step in the analysis is the reconstruction of $\D{0} $ and $\D{+} $ 
mesons. In case of $\D{0} $ a low momentum pion $\pi_{sl}$ 
($p^* = 40 \, {\mathrm MeV}/c$) with the correct charge correlation with the 
meson is searched for the $\D{*+} $ reconstruction.
The most relevant characteristics of the DELPHI detector for this part 
of the analysis are the particle identification provided by the RICH and 
the specific energy loss $dE/dx$ measured by the TPC~\cite{delphi-nim}. 
These informations are combined into a pion veto used in the $\D{0/+} $ 
channels to reduce the combinatorial background which mainly consists of 
misidentified pions.
 
\begin{figure}[t]      % in second brace, h=here, t=top, b=bottom      
\centerline{
\begin{tabular}{cc}
\epsfxsize 2.0 truein \epsfbox{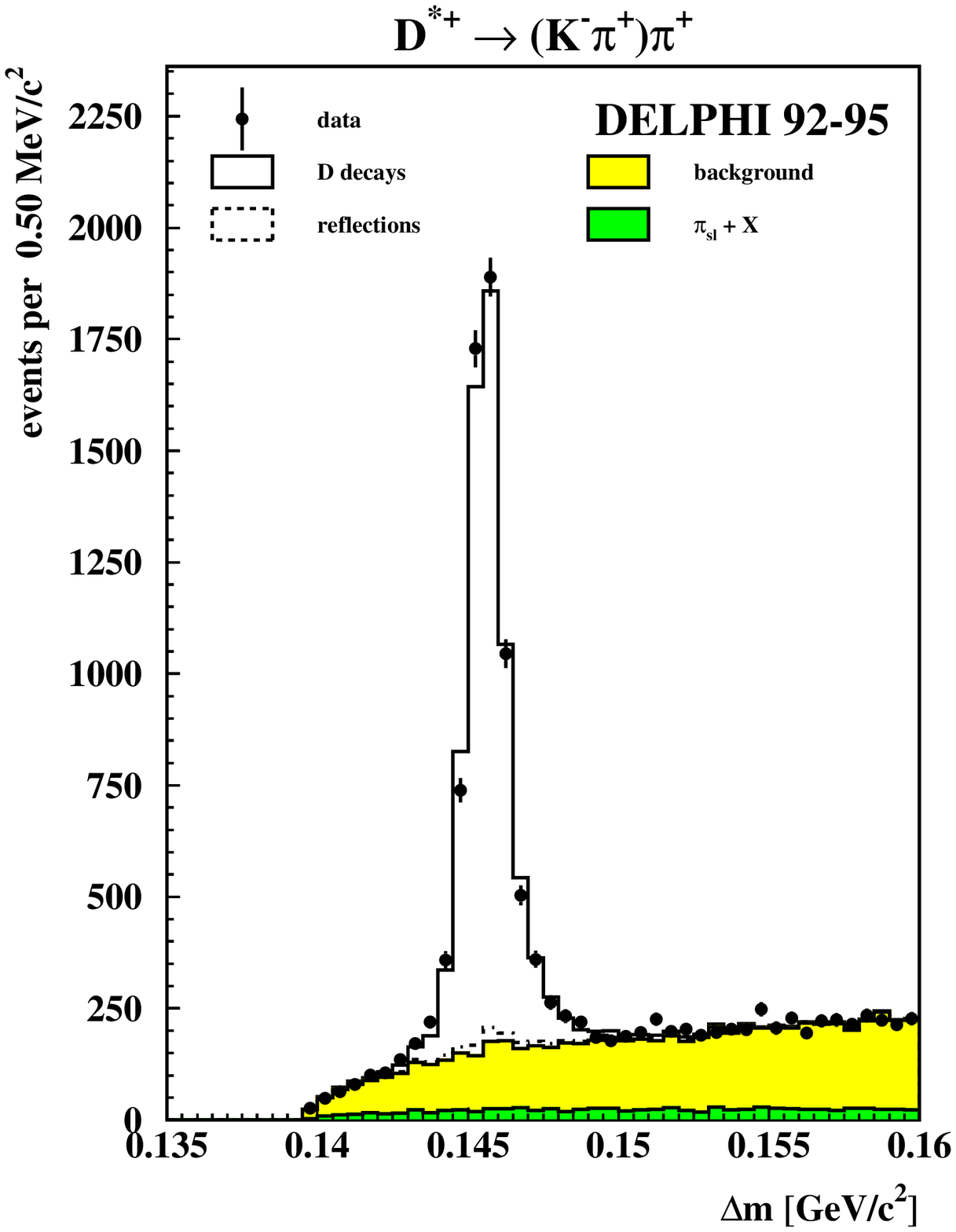}   & \epsfxsize 2.0 truein \epsfbox{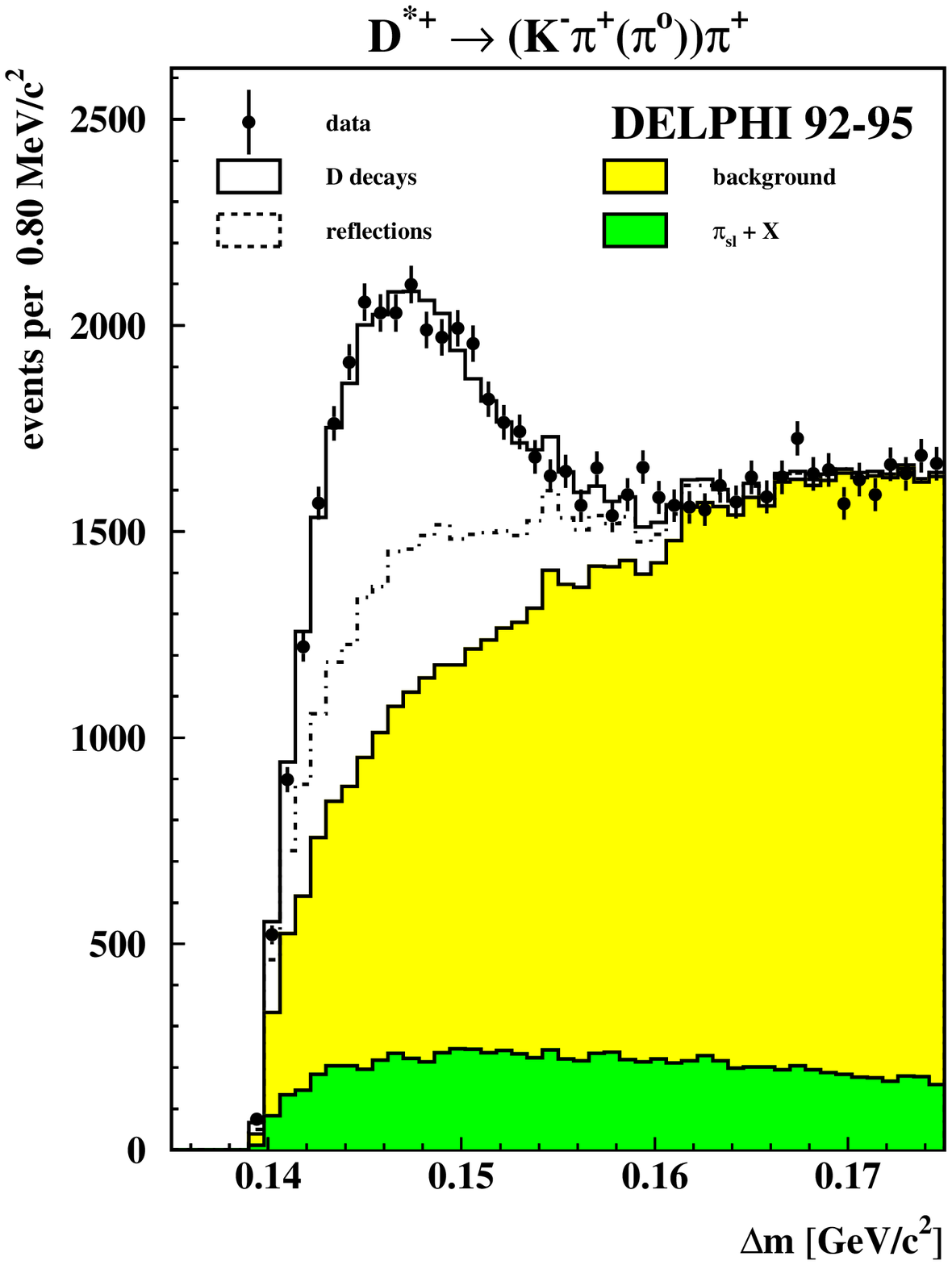}   \\
\end{tabular}}
\vskip .5 cm
\caption[]{
\label{fig:D1}
\small The mass difference $\Delta m$ spectra for $\D{*+} \ra (\K{-}
\pi^+) \pi^+_{sl}$ and  $\D{*+} \ra (\K{-} \pi^+ (\pi^0) ) \pi^+_{sl}$
decay modes for data and simulation.}
\end{figure}

Further reduction of the background is achieved with cuts in the helicity 
angle of the kaon candidate with respect the $\D{0/+} $ flight direction and 
in the 3D decay length of the meson.
These cuts are applied in function of the scaled energy of the meson 
$X_E(D) = 2 E_D/\sqrt{s} $ to account for the energy spectrum of charged 
particles in hadronic Z decays, peaked at  low momentum.
Finally candidates are selected either in a $\Delta m = m_{\D{*} } - 
m_{\D{0} }$ region for $\D{*+} $ channels, or in a mass interval for a 
$\D{0/+} $ decay.
The range of $\Delta m$ values goes from 160 MeV/$c^2$ to 250 MeV/$c^2$ 
accordingly to the invisible energy in the final state, while for $\D{+/0} $ 
channel a mass region $\pm 200$ MeV/$c^2$ around the 
nominal mass is selected. The reconstructed mass spectra for two channels are 
shown in fig.~\ref{fig:D1}

\begin{figure}[b]      % in second brace, h=here, t=top, b=bottom      
\centerline{
\begin{tabular}{cc}
\epsfxsize 2.0 truein \epsfbox{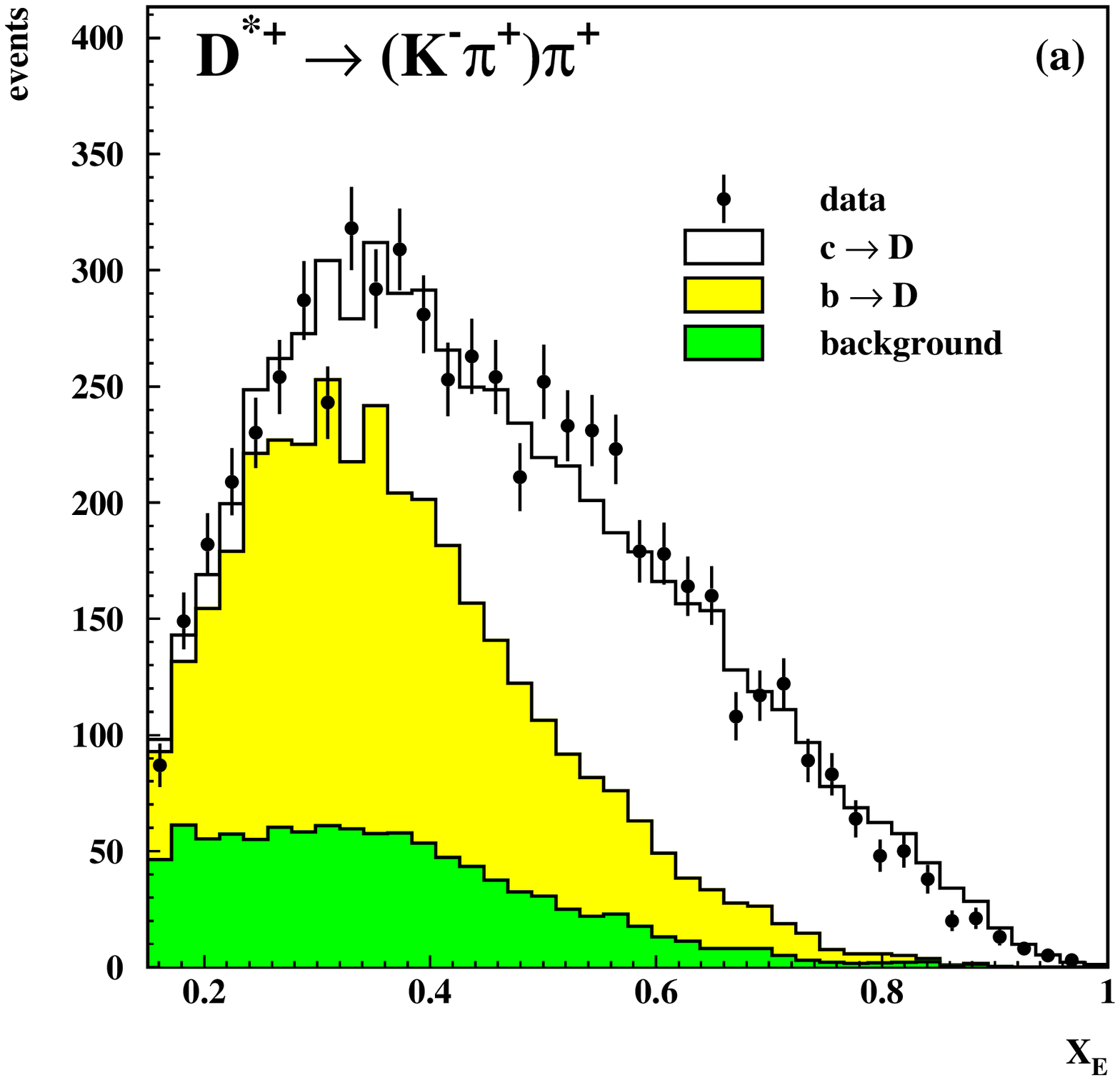}   & \epsfxsize 2.0 truein \epsfbox{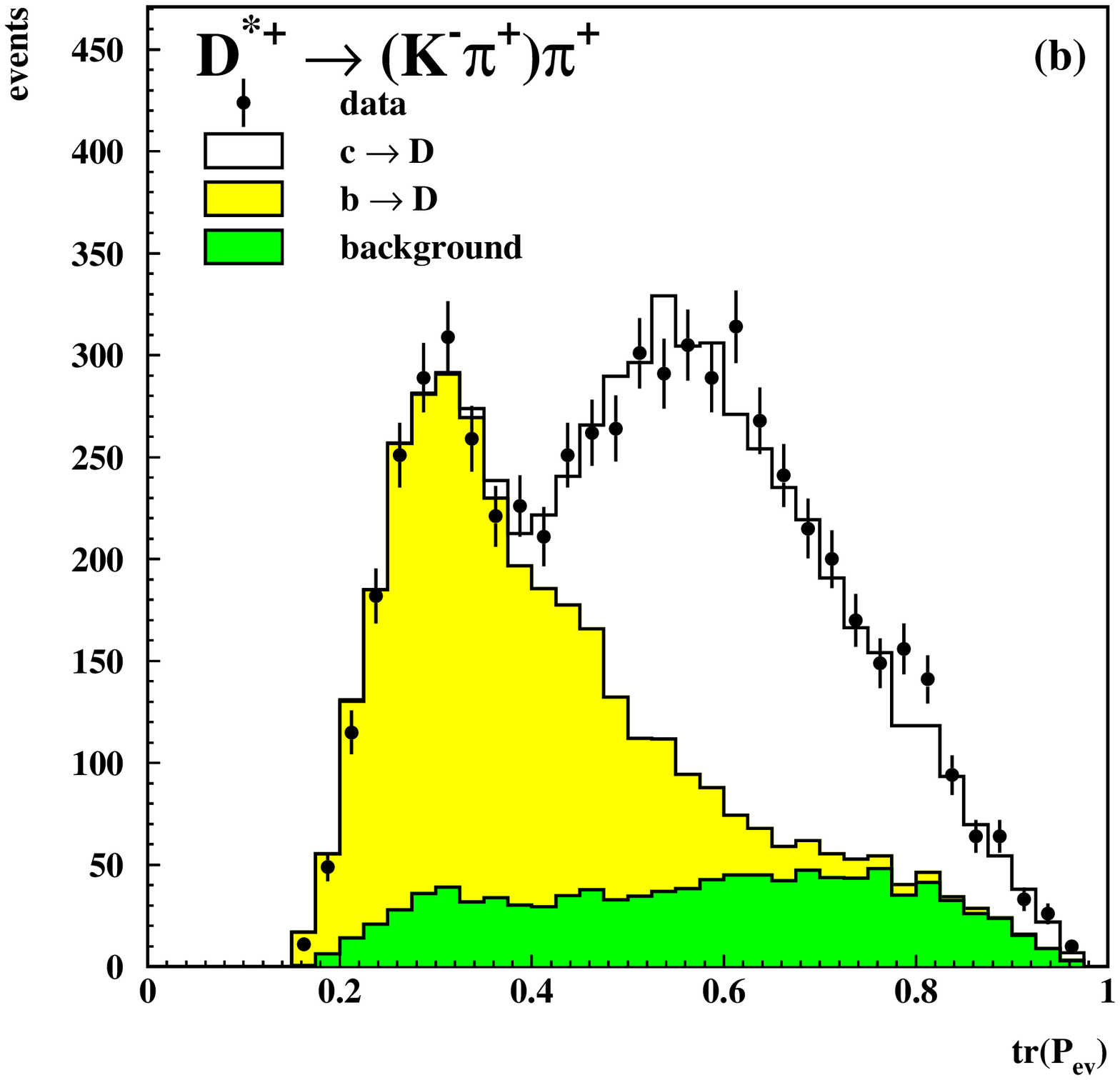}   \\
\end{tabular}}
\vskip .5 cm
\caption[]{
\label{fig:D2}
\small The distributions of $X_E(D)$ and $b$-tag probability for the
signal region for $\D{*+} \ra (\K{-} \pi^+) \pi^+_{sl}$ decay for data
and simulation.}
\end{figure}

To extract $\Afbp{c} $ and $\Afbp{b} $ the contributions of $\D{} $ mesons 
from $\ff{c} $ and $\ff{b} $ events should be separated. This is achieved 
fitting together $\Afbp{c} $ and $\Afbp{b} $ in bins of  $\cos \theta_T$, 
scaled momentum $X_E(D)$ and $b$-tag variable ${\cal P}_{ev} $ 
(fig.~\ref{fig:D2}).

In fig.~\ref{fig:D1} one can notice that, in addition to the pure
combinatorial background, in the signal region there are two other components: 
Partially reconstructed $\D{*+} $ mesons and reflections from other decays mode.
Because of the charge correlation with the primary quark, these two
categories have to be treated like signal in the fit. 

Finally in case of $\ff{b} $ events an effective mixing is introduced
to take into account the $\B{0} \bar{\B{0} } $ mixing for  $\B{} \ra \D{} $ 
produced mesons and the so-called ``upper vertex'' production of the charm in 
$b \ra \D{} \bar{\D{} }$ decays. 
This was computed from data collected at the $\Upsilon(4s)$ resonance by the 
CLEO and the ARGUS collaborations. The values obtained are $\chi_{eff}=0.222 
\pm 0.033$ for $\D{+} $ and $\D{*+} $ modes and $\chi_{eff}=0.170 \pm 0.030 $ 
for the  $\D{0} $ channels, both different from the average $\overline{\chi} = 
0.1214 \pm 0.0043$ measured at LEP for $b$-hadrons.

At the Z peak about $62 \times 10^3$ $\D{} $ decays were reconstructed and the 
corresponding value of the asymmetries are $\Afb{c} = 0.0659 \pm
0.0094$ (stat) and $\Afb{b} = 0.0762 \pm 0.0194$ (stat).
The main contributions to the systematic error are listed in the bottom part 
of table~\ref{tab:syst}. Besides the limited MC statistics the main 
contributions are related to the fit method itself and to the residual 
asymmetry of $\D{} $ mesons from pure combinatorial background in $\ff{c} $ 
and $\ff{b} $ events.
%\pagebreak

\section{LEP combined results.}

\begin{figure}[b]      % in second brace, h=here, t=top, b=bottom      
\centerline{
\begin{tabular}{lr}
\epsfxsize 3.2 truein \epsfbox{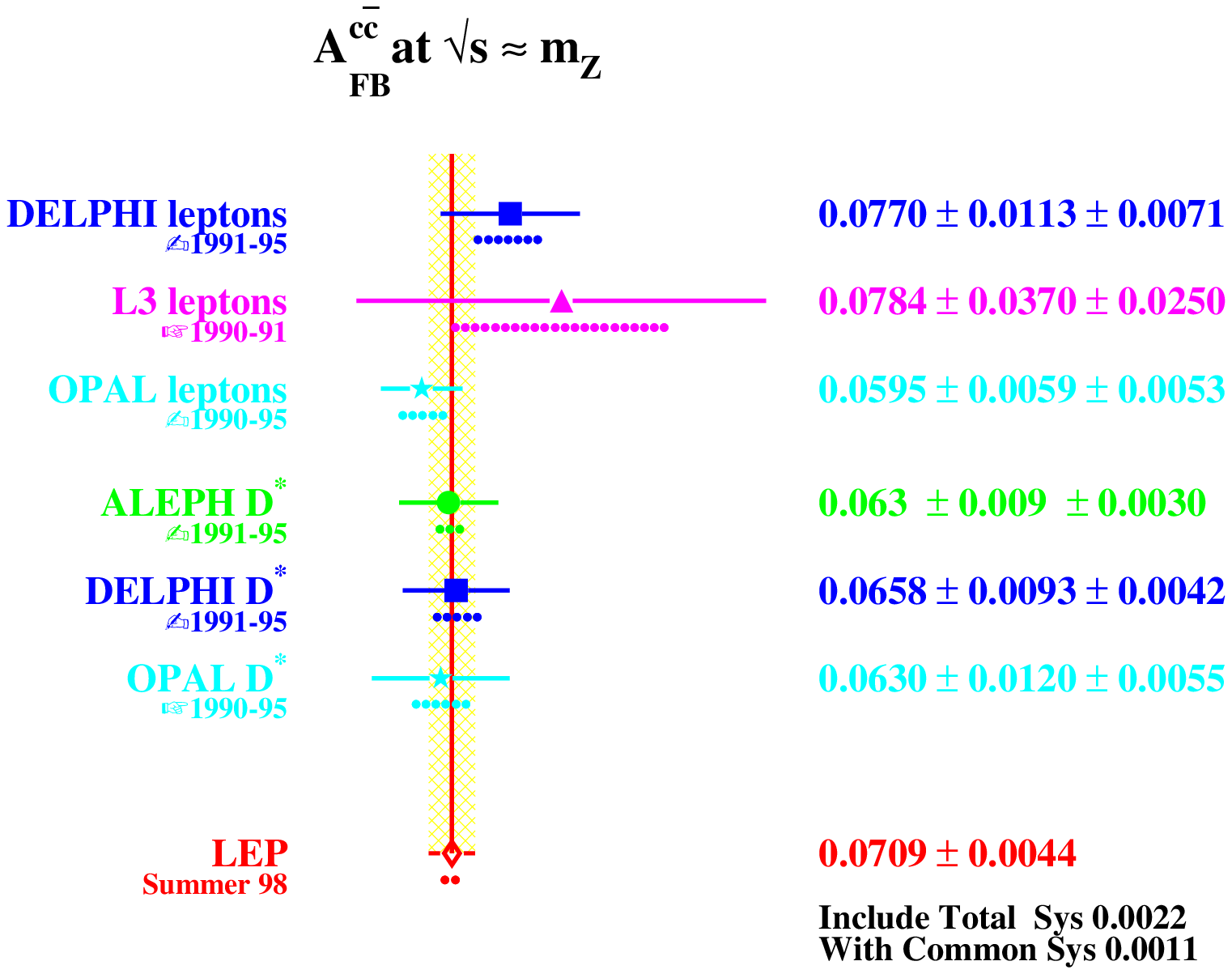}   & \epsfxsize 3.2 truein \epsfbox{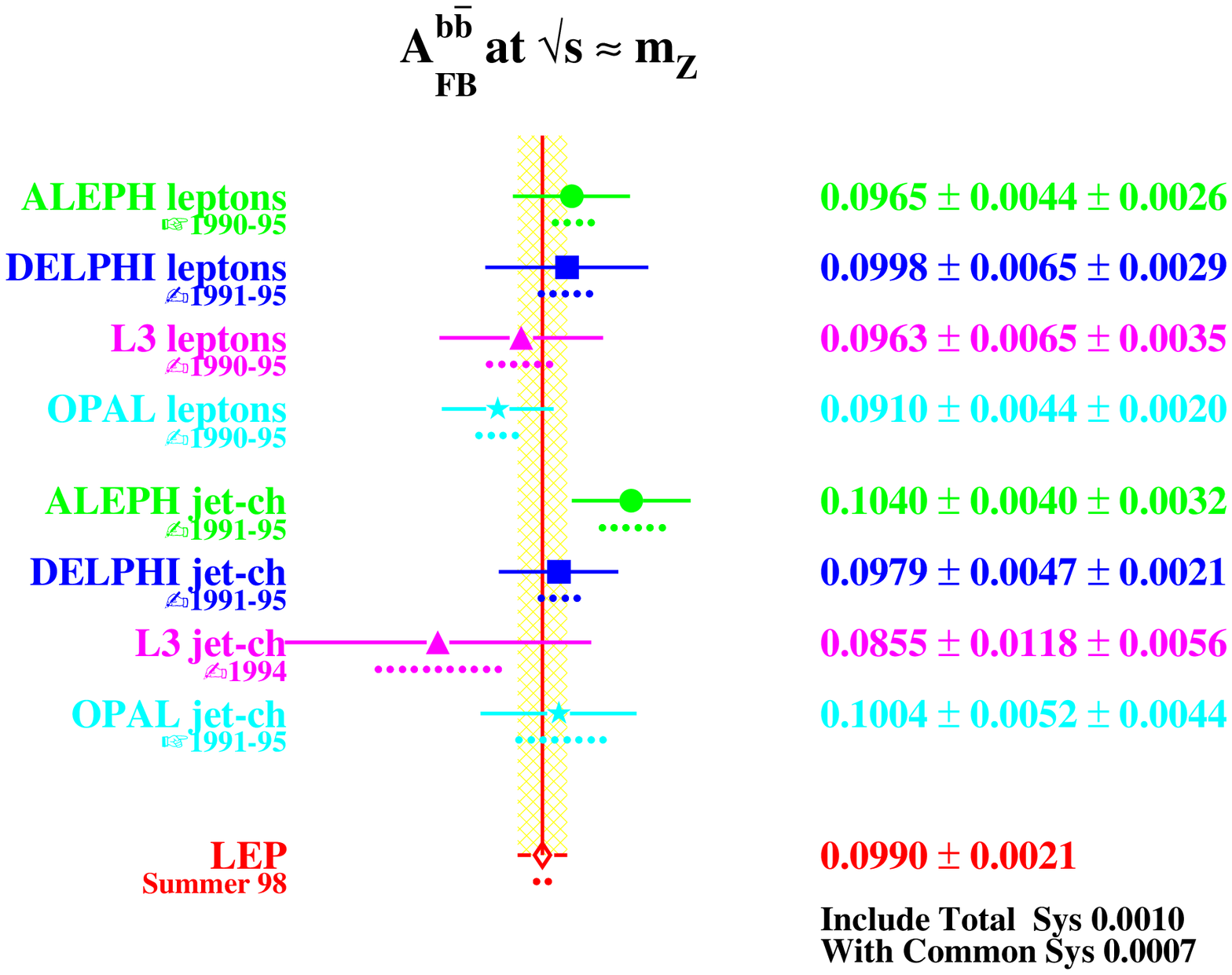} \\
\end{tabular}}
%\vskip -.2 cm
\caption[]{
\label{fig:LEP1}
\small Results for $\Afbp{c} $ and $\Afbp{b} $ shown at the ICHEP 98 Vancouver Conference.}
\end{figure}

The currently available determinations of  $\Afbp{c} $ and
$\Afbp{b} $ from LEP experiments are shown in figure~\ref{fig:LEP1}. The
precision on the measurement of $\Afbp{c} $ is currently 7 \% and the average
is dominated by the measurements performed with the $\D{} $ meson tag.
On the contrary, it should be noted that the most  precise single
determination comes from the lepton analysis tag from OPAL. This
indicates that some improvements in the measurement of $\Afb{c} $ can
still be achieved. The precision on the measurement of $\Afb{b} $ is
currently 2 \% with equal weights from jet charge and lepton tag
analyses. The impact of these measurements on the determination of
$\sWe{\ell} $ is 

\begin{eqnarray*}
\sWe{\ell } = 0.2322 \pm 0.0010   \,\, \mathrm{for} \,\, \Afbp{c} \\
\sWe{\ell } = 0.23225 \pm 0.00038 \,\, \mathrm{for} \,\, \Afbp{b} 
\end{eqnarray*}

$\Afbp{b} $ provides together $\Alr $ from SLD the most precise value available 
at the moment. On the contrary the two measurements are more than 2 sigmas
apart from each other.

%% ONLY POSTSCRIPT FILES CAN BE INCLUDED

%%%%%%%%%%%%%%%%%%

%%%Sample Table; Please follow this format

\end{document}